# Exploring the Coexistence of Spin States in [Fe(tpy-Ph)$_2$]$^{2+}$ Complexes on Au(111) using ab initio calculations

Naveen K. Dandu,[1,2][*] Alex Taekyung Lee,[1,2][*] Sergio Ulloa,[3] Larry Curtiss,[2] Saw Wai Hla,[3,4] and Anh T. Ngo[1,2]

[1]Department of Chemical Engineering, University of Illinois-Chicago, Illinois, 60608, USA

[2]Materials Science Division, Argonne National Laboratory, Lemont, IL 60439, USA

[3]Nanoscale and Quantum Phenomena Institute and Department of Physics and Astronomy, Ohio University, Athens, OH 45701, USA

[4]Nanoscience and Technology Division Argonne National laboratory, Lemont, IL 60439, USA

**Abstract:**

In this work, we systematically study the electronic structure and stability of spin states of the [Fe-(tpy-ph)$_2$]$^{2+}$ molecule in both gas phase and on a Au(111) substrate using density functional theory +U (DFT+U) calculations. We find that the stability of the Fe$^{2+}$ ion's spin states is significantly influenced by the Hubbard U parameter. In the gas phase, the low-spin (LS, S=0) state is found to be energetically favorable for U(Fe) ≤ 3 eV, whereas the high-spin (HS, S=2) state is stabilized for U(Fe) > 3 eV. Interaction with the Au(111) substrate is found to elevate the critical U for the spin-state transition to 3.5 eV. Additionally, we perform L-edge X-ray absorption spectroscopy (XAS) calculations based on time-dependent DFT (TD-DFT) for both HS and LS states. The calculated XAS suggests that the HS state more closely aligns with the experimental observations, indicating the potential coexistence of the HS state as the initial state during the X-ray excitation process. These findings enrich our understanding of spin-state dynamics in [Fe(tpy-Ph)$_2$]$^{2+}$.

*Naveen Dandu and Alex Taekyung Lee contributed equally to this work.

# I. Introduction

Since its discovery in the 1980s, spintronics has emerged as a pivotal field in the development of nanoelectronic devices, primarily due to its low power consumption and enhanced memory and processing capabilities.[1,2] Among these, especially, organic semiconductors (OSCs) have garnered particular interest for their superior spin transport characteristics at room temperature,[3] a feature that has garnered increased attention. This interest is further fueled by the ease of functionalizing OSCs and their distinctive interfacial properties when interfaced with electrodes, paving the way for diverse applications. Concurrently, single molecule magnets (SMMs) have demonstrated promising potential in areas such as data storage, quantum systems, and spintronics, offering an appealing alternative to traditional materials.[4–6] Notably, mononuclear transition metal complexes from the first row exhibit intriguing magnetic properties, distinguishing themselves within the realm of SMMs for their unique contributions to spintronics.[4,7,8]

One of the intriguing phenomena observed in molecular magnets is the spin crossover (also known as spin transition), which entails transitions between different spin states of transition metal (TM) ions. These ions can exhibit multiple spin states due to crystal field splitting, influenced by the local environment surrounding the TM ion. Spin-state transitions can be induced through various mechanisms, including thermal excitation,[9–12] photoexcitation,[13,14] or epitaxial strain in solids.[15–17] Among molecular magnets, the $Fe^{2+}$ ion in $[Fe(L)_2]^{2+}$ (where L = tpy, bpy, etc) are renowned for their spin-state transitions, as demonstrated in various studies.[14,18–22] The ground state of these complexes is primarily a low-spin (LS) state with S=0 ($t_{2g}^6 e_g^0$), though the stability of this state significantly varies with the chosen exchange-correlation functional.[20–22] This variation highlights the critical role of exchange interactions and electron-electron correlations in determining the stability of spin states. For instance, in $[Fe(bpy)_2]^{2+}$, the high-spin (HS) state with S=2 ($t_{2g}^4 e_g^2$) proves to be more stable than the LS state by 0.134 eV under the B3LYP hybrid functional, in contrast to the LS state being the ground state when using BP86 and TPSSH functionals.[22] The stability of the LS state hinges on the magnitude of $t_{2g}$-$e_g$ splitting (as illustrated in Fig. 2), which is significantly influenced by the electron correlation in the Fe d orbitals. Consequently, understanding the complexities of electron correlation in $[Fe(L)_2]^{2+}$ on Au (111) particularly through the lens of Hubbard U within DFT+U studies, is crucial for a deeper insight into the spin state stabilities in these complexes.

Several studies have utilized X-ray spectroscopy techniques, including X-ray absorption spectroscopy (XAS)[13] and X-ray emission spectroscopy (XES) alongside resonant X-ray emission spectroscopy (RXES),[14] to investigate the electronic and magnetic structure of *[Fe(tpy-Ph)$_2$]$^{2+}$*. Recently, a new X-ray technique known as synchrotron X-ray scanning tunnelibng microscopy (SX-STM) has been used to investigate the near-edge X-ray absorption fine structure (NEXAFS) of *[Fe(tpy-Ph)$_2$]$^{2+}$*.[23]

In this work, we systematically investigate the electronic structure and spin-state stabilities of the free *[Fe(tpy-Ph)$_2$]$^{2+}$* molecule (Fig. 1a) and its Au(111) substrate-bound form using DFT+U calculations. We demonstrate that low-spin states prevail at lower U values, while increased electron correlation in Fe d orbitals favors high-spin states. Importantly, we find that the interaction with the Au substrate increases the critical U necessary for spin-state transitions. Additionally, we perform XAS simulation for both spin states using time-dependent DFT. In the following sections, the methods used in our computational approach followed by details of the obtained results are described.

## II. Computational Methods

### 1. Geometry optimization using DFT+U

All the structures are optimized utilizing plane wave basis sets and projector-augmented wave (PAW) pseudopotentials implemented in VASP code.[23] For the geometric relaxations, exchange-correlation functionals were used within the generalized gradient approximation (GGA) of Perdew-Burke-Ernzerhof (PBE) with U corrections.[24] The cutoff energy for the plane-wave basis was 600 eV, which was tested and applied to all supercells. The convergence criterion of the total energy was set to be within $1 \times 10^{-6}$ eV within the K-point integration. The Brillouin zone was sampled at Γ-point only, and all the geometries were optimized until the residual forces per atom were less than 0.001 eV/Å.

Calculations were conducted for both *[Fe(tpy-Ph)$_2$]$^{2+}$* and *[Fe(tpy-Ph)$_2$]$^{2+}$* adsorbed on three layers of the Au (111) surface (refer to Figure 1). To maintain the structural integrity of the model and prevent bending of the Au slab, a three-layer Au (111) slab was used and only the top two layers of the slab were relaxed. The Au layer's dimensions were set to 20.19Å × 29.98 Å to ensure sufficient separation between *[Fe(tpy-Ph)₂]²⁺* molecules. The minimum distance between [Fe(tpy-ph)₂]²⁺ molecules in adjacent supercells was maintained at 15 Å, which effectively minimizes inter-supercell interactions. Additionally, a vacuum layer of 15 Å was utilized in the slab geometry to simulate isolation. Given the significance of non-covalent interactions on the complex surface, we incorporated the van der Waals D3 functional[25] in our calculations.

## *2. XAS using TD-DFT*

All calculations were performed using the ORCA software of version 5.0.3.2.[26] The X-Ray absorption spectra (XAS) were calculated using density functional theory (DFT) combined with Restricted Open Shell Configuration Interaction with Singles (DFT-ROCIS) method.[27] The B3LYP functional[28–30] was used with the Ahlrich polarized def2-TZVP(-f) basis sets[31,32] together with the auxiliary def2/J functions to accelerate the calculations.[33,34] Scaling parameters of c1=0.18, c2=0.20 and c3=0.40 were used, which is suitable for the B3LYP functional.[35] The excitation window was constructed specifying one donor space corresponding to the excitation donor orbitals, Fe '*2p*', and an acceptor orbital space that corresponds to all singly occupied states and the entire virtual orbital space. A total of 100 roots were requested to guarantee the saturation of all excitations. The data files for generating plots were produced using a utility called Orca_mapspc, which was integrated into the Orca program suite.

Spin-orbit coupling (SOC): The SOC is calculated using the QDPT based on the non-relativistic roots. The CI procedure leads to excited state wavefunctions of the form $|\Psi_{ROCIS}^{SS}\rangle = \sum_\mu C_{\mu_I}|\Phi_\mu^{SS}\rangle$. The upper indices SS denote a many-particle wavefunction with spin quantum number S and spin projection quantum number M = S. Since the BO Hamiltonian is spin free, only one member of a given spin-multiplet (e.g. that with M = S) needs to be calculated. For the treatment of the SOC, all $|\Psi_I^{SM}\rangle$ are required ('I' denotes a given state obtained in the first step of the procedure). Matrix elements over the $|\Psi_I^{SM}\rangle$ functions are readily generated using the Wigner–Eckart theorem, since all (2S + 1) members of the multiplet share the same spatial part of the following wavefunction[36]:

$$\langle\Psi_I^{SM}|\hat{H}_{BO} + \hat{H}_{SOC}|\Psi_J^{S'M'}\rangle = \delta_{IJ}\delta_{SS'}\delta_{MM'}E_I^{(S)} + \langle\Psi_I^{SM}|\hat{H}_{SOC}|\Psi_J^{S'M'}\rangle$$

The above equation yields spin-orbit coupled eigenstates and the corresponding eigen energy values. Finally, the SOC operator is approximated by the effective one-electron spin-orbit field operator.

## III. Results and Discussion

The structure of the derivative of a *[Fe(tpy-Ph)$_2$]$^{2+}$* molecule on a Au(111) surface is depicted in Figure 1, showcasing the terpyridine ligands renowned for their strong binding affinity to transition metals, which encase the central Fe ion. The Fe atom is coordinated by six nearest neighbor N atoms, forming an Fe-N$_6$ octahedron, as shown in Figure 2(a). This octahedral coordination leads to the splitting of Fe d orbitals into lower energy *t$_{2g}$* and higher energy e$_g$ orbitals. The Fe-N6 octahedron demonstrates tetragonal (D$_{2d}$) symmetry and exhibits Jahn-Teller (JT) distortion, leading to a noticeable difference in bond lengths. Specifically, as presented in Figure 3, the in-plane Fe-N bonds are longer than those out-of-plane. Consequently, this symmetry causes the e$_g$ orbitals to further split into $d_{z^2}$ (at lower energy) and $d_{x^2-y^2}$ (at higher energy) orbitals. Similarly, the t$_{2g}$ states divide into doubly degenerate d$_{xz}$/d$_{yz}$ orbitals (at lower energy) and a singular d$_{xy}$ orbital (at higher energy), as schematically illustrated in Figure 2(b).

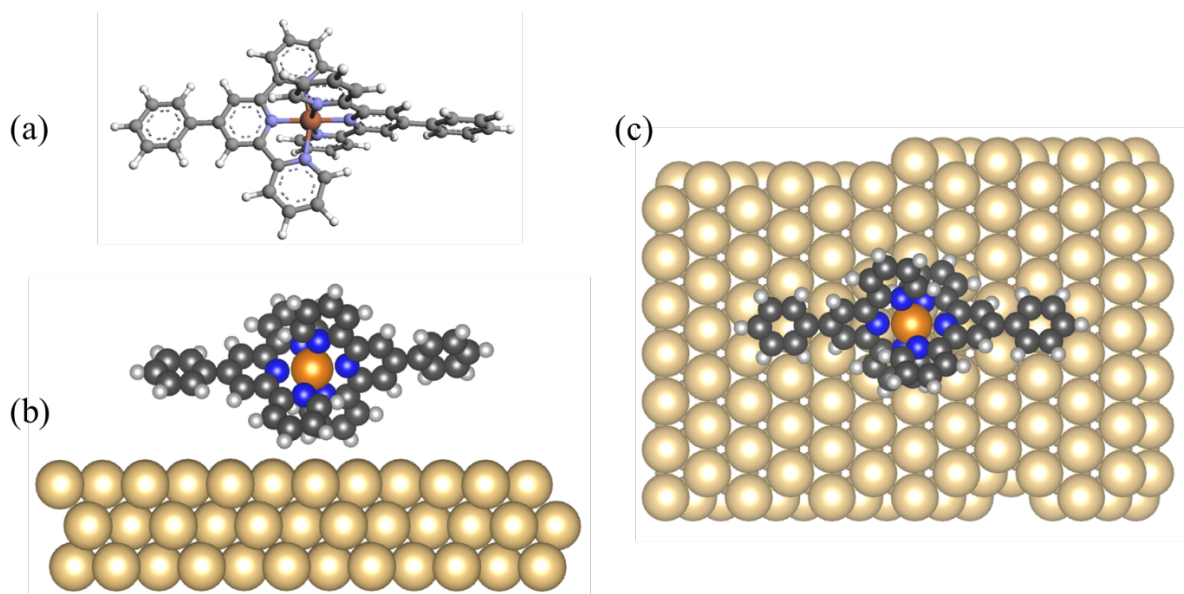

**Fig 1: Structure and adsorption geometry.** *(a) Molecular structure of [Fe(tpy-Ph)$_2$]$^{2+}$, (b) Relaxed geometry of [Fe(tpy-Ph)$_2$]$^{2+}$on a Au(111) support in side view and (c) top view. Black: C atoms, grey: H atoms, blue: N atoms, and brown: Fe atom.*

The $Fe^{2+}$ ion can exhibit three distinct spin states: (i) a low-spin (LS) state with S=0 [Figure 4(a)], characterized by a fully occupied $t_{2g}$ ($t_{2g}^6$) and completely empty eg ($e_g^0$) orbitals; (ii) an intermediate-spin (IS) state with $S$=1 [Figure 4(b)], where four spin-up $d$ orbitals are filled; three spin-up $t_{2g}$ and one $e_g$—alongside two spin-down $t_{2g}$ orbitals, resulting in S=1 (Note: in this state, the spin-up $d_{z^2}$ orbital is filled due to its lower energy, a consequence of Jahn-Teller (JT) distortion under $D_{2d}$ symmetry and the $d_{x^2-y^2}$ orbital remains unoccupied); (iii) a high-spin (HS) state with S=2 [Figure 4(c)], marked by all spin-up d orbitals being filled, and one spin-down $t_{2g}$ orbital partially occupied. Given the $D_{2d}$ symmetry, both spin-down $d_{xz}$ and $d_{yz}$ orbitals are half-filled, while the spin-down $d_{xy}$ orbital remains empty due to JT distortion effects. We note that all spin states are insulating within GGA+U (U=3), as shown in Figure 4, with the HOMO-LUMO gap being more constricted in the LS state than in the HS state.

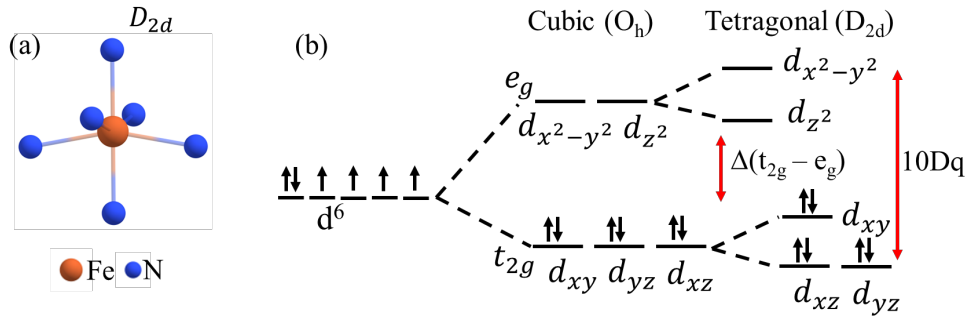

**Fig 2:** *(a) depiction of Oh symmetry of the the Fe-N6 octahedron environment and (b) crystal field orbital splitting depiction of $Fe^{2+}$ ion. In this figure, N atoms are colored in blue and Fe in orange.*

In Fig. 5(a), the calculated total energies of the low-spin (LS), intermediate-spin (IS), and high-spin (HS) states of the *[Fe(tpy-Ph)$_2$]$^{2+}$* complex in the gas phase are plotted as functions of U(Fe). To quantify the relative stability of these spin states, the energy difference, ΔE = E[spin state] - E[LS], is utilized. The LS state remains the energetically favored ground state for U(Fe) values up to 3 eV. Beyond this threshold, the HS state becomes the ground state for U(Fe) > 3 eV. At U=3 eV, the calculated energy difference between the HS and LS states, $\Delta E_{HS-LS}$ = E[HS] - E[LS], is determined to be 9.52 meV. The stabilization of the LS state is attributed to the considerable $t_{2g}$-$e_g$ orbital separation, which is pronounced when the Fe-N bond length is minimized, enhancing the Fe d – N p orbital interactions. This stabilizing effect diminishes with an increase in U, corresponding to the elongation of the Fe-N bond and increased localization of Fe d orbitals at higher U values, as shown in Fig. 3. Conversely, the stability of HS state is bolstered by the augmented spin-splitting of the d orbitals, an effect that intensifies with the increment of U values within the DFT+U framework, thereby making the HS state energetically preferable at elevated U values.

Additionally, we note that previous studies have examined the energy difference between low-spin (LS) and high-spin (HS) states ($\Delta E_{HS-LS}$) of $Fe^{2+}$ ions in Fe(II) molecular complexes using the PBE+U method with U values of 0, 2, 4, 6, and 8 eV, and compared these results with the CASPT2/CC method.[37] For $[Fe(NH_3)_6]^{2+}$ and $[Fe(NCH)_6]^{2+}$, U values of 2 to 4 eV provide the best match with the CASPT2/CC method, while the HS state is more stable for both molecules within the CASPT2/CC framework.[37]

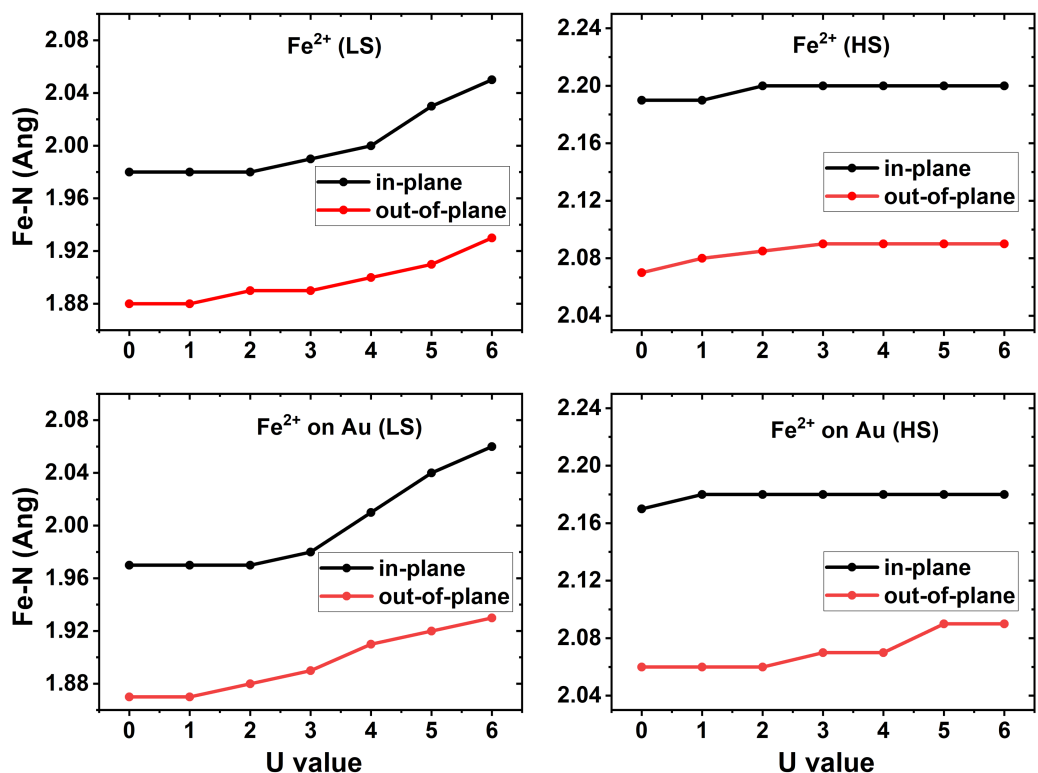

**Fig 3**: *Fe-N bond lengths in [Fe(tpy-ph)₂]²⁺ for (a) the LS state and (b) the HS state in the gas phase, and in [Fe(tpy-ph)₂]²⁺ on a Au substrate for (c) the LS state and (d) the HS state.*

Given the predominantly van der Waals interaction between *[Fe(tpy-Ph)$_2$]$^{2+}$* and Au, the electronic structure of Fe d states in both the gas phase of *[Fe(tpy-Ph)$_2$]$^{2+}$* and on Au remain similar within the GGA+U framework, as shown in Figures 4 and 6. The notable distinction lies in the Fermi level (EF) position. The Fermi level (E$_F$) for *[Fe(tpy-Ph)$_2$]$^{2+}$* on Au is determined by the orbitals of Au.

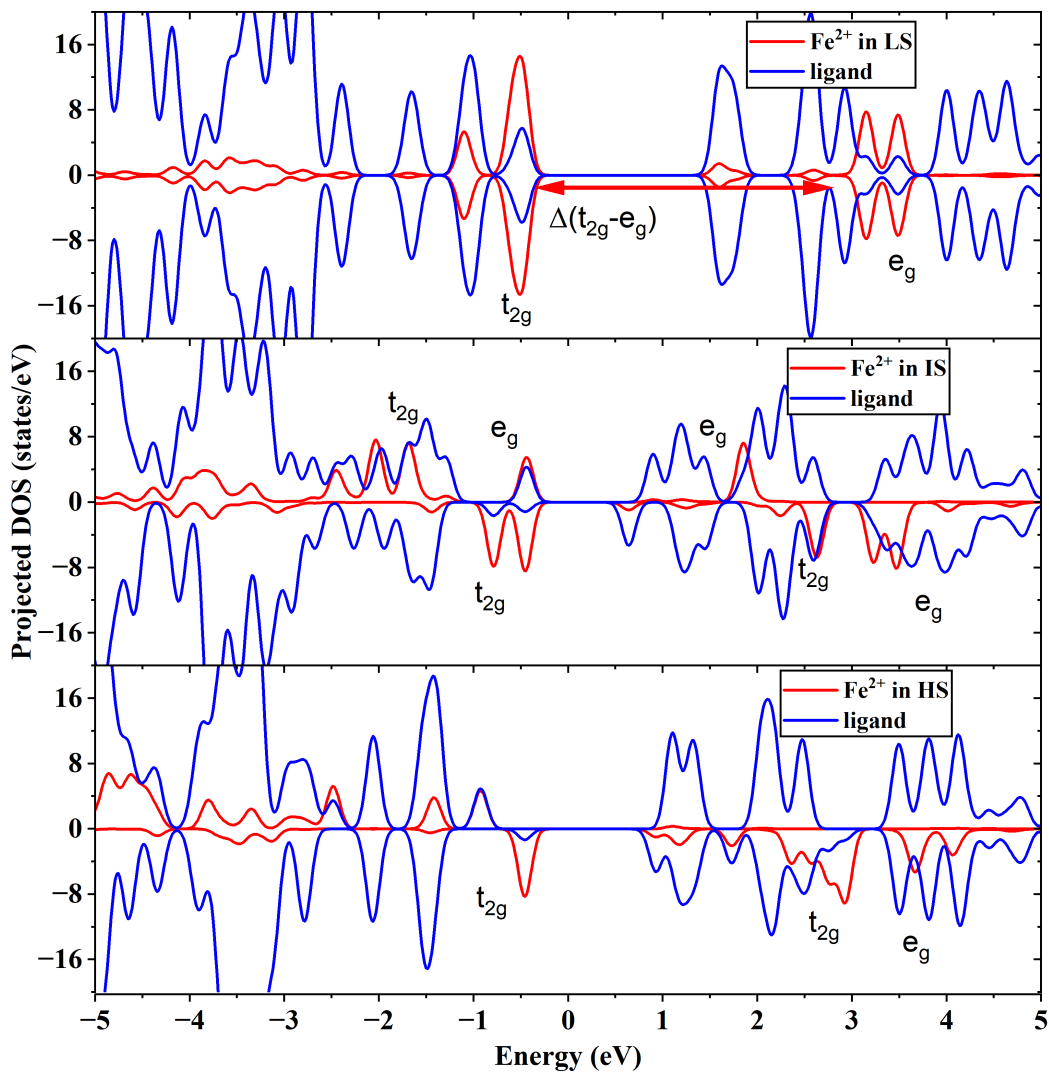

**Fig 4:** *Projected density of states for Fe 'd' orbitals of [Fe(tpy-Ph)$_2$]$^{2+}$ complex in their LS, IS and HS states.*

To investigate the impact of the *[Fe(tpy-Ph)$_2$]$^{2+}$* on a Au substrate interaction on spin-state stability, Figure 5(b) shows the calculated energy differences (ΔE) of spin states for [Fe(tpy-

ph)2]2+ on Au(111) , plotted as a function of the Hubbard U parameter. The results closely match those observed in the gas phase, aligning with the expected outcome due to the van der Waals-dominated molecule-Au interaction: the low-spin (LS) state is energetically preferred for U values less than 3.5 eV, while the high-spin (HS) state emerges as the ground state when U exceeds 3.5 eV. This identifies a critical U value for the HS-LS transition at 3.5 eV, slightly increased from that in the gas phase, underlining the subtle yet significant influence of the Au substrate on spin state stability. Further, we note changes in Fe-N bond lengths due to the molecule-Au interaction, as shown in Figure 3, which could affect the stability of the spin states. Specifically, the in-plane Fe-N bond lengths within the HS phase of *[Fe(tpy-Ph)$_2$]$^{2+}$* on Au are reduced compared to those in the gas phase, as demonstrated in Figures 3(b) and (d).

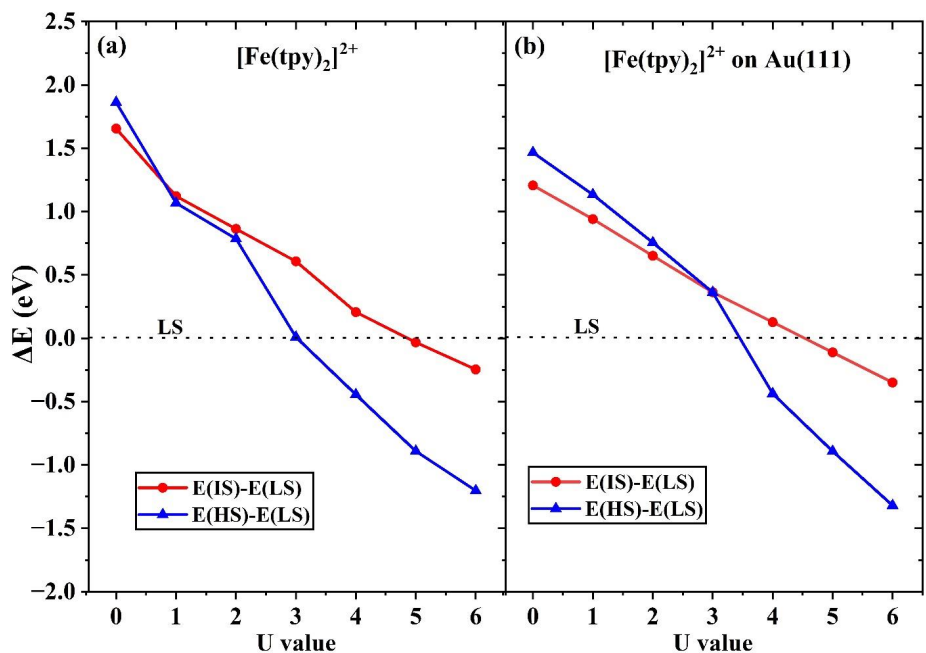

***Fig 5***: *Values of ΔE = E[HS or IS] – E[LS] at varying U value from 0 to 6. Here HS = high-spin (S=2), IS = intermediate-spin (S=1) and LS = low-spin (S=0) corresponding to (a) [Fe(tpy-Ph)$_2$]$^{2+}$ and (b) [Fe(tpy-Ph)$_2$]$^{2+}$on Au(111).*

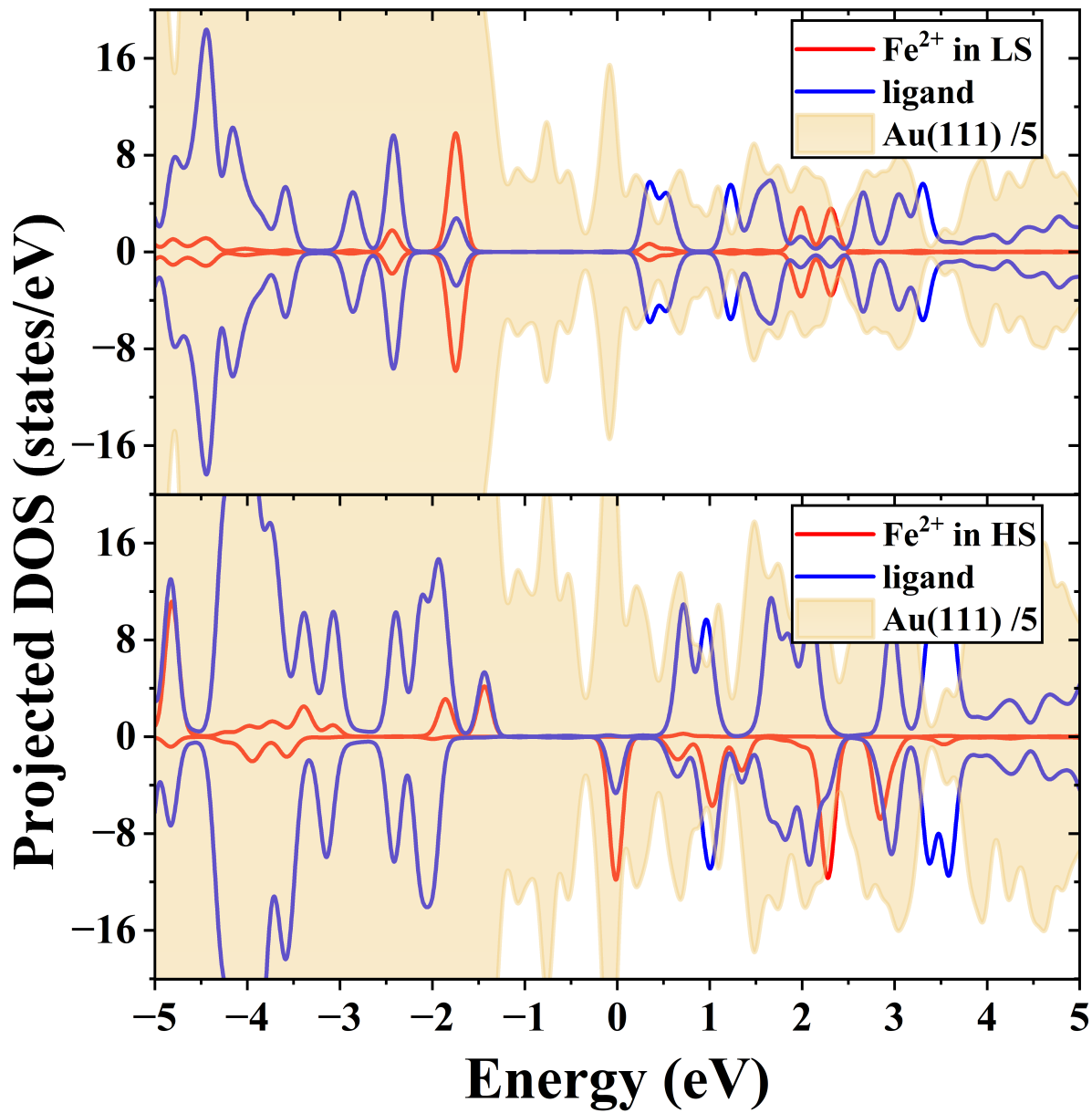

*Fig 6.* *Projected density of states of [Fe(tpy-Ph)$_2$]$^{2+}$ on Au(111) surface in their LS and HS states.*

Next, we consider the X-ray absorption spectra (XAS) of *[Fe(tpy-Ph)$_2$]$^{2+}$*. Building on our examination of the electronic structure of *[Fe(tpy-Ph)$_2$]$^{2+}$* on a Au substrate with SX-STM and analysis of the low-spin (LS) state using DFT+U, as detailed in ref [38], we extend our inquiry to the absorption properties of this complex. Despite the prevalent consideration of the LS state as the ground state for the Fe$^{2+}$ ion in *[Fe(tpy-Ph)$_2$]$^{2+}$* molecules, as referenced in various studies,[14,18–22] experimental conditions may enable the coexistence of multiple spin states, with their stability dependent on the Hubbard U parameter. Various factors, such as thermal and photoexcitation or an intensified effective correlation from the local environment of Fe ion, could promote the high-spin (HS) state to the ground state. Thus, it is imperative to consider both HS and LS phases when analyzing XAS, acknowledging the excitation-induced potential for spin state transitions.

As XAS method has limitations in terms of the size of the system, we calculated only for the isolated molecules. From the plotted density of states shown in figures 4 and 6, it can be seen that there are no remarkable changes in the characteristic peaks of the unoccupied states with or without Au(111), in comparison. Henceforth, we assume that XAS spectra will not be effected in the presence of a Au(111) surface. This will also reduce a substantial amount of the computational cost.

Utilizing time-dependent density functional theory (TD-DFT) for the calculation of the L3-edge X-ray absorption spectra (XAS) of *[Fe(tpy-Ph)$_2$]$^{2+}$*, we model the electron excitation from Fe 2p to 3d states. This comprehensive analysis includes both low-spin (LS) and high-spin (HS) states in the gas phase of *[Fe(tpy-Ph)$_2$]$^{2+}$* and juxtaposes these results with experimental synchrotron X-ray scanning tunneling microscopy (SX-STM) data, as illustrated in Figure 7.[38] The experimental spectrum, which ranges from 706 eV to 712 eV, reveals six peaks (denoted as i-vi in Figure 7). In this context, the HS phase replicates all six observed peaks (i-vi), consistent with the experimental data, while the LS phase manifests only five peaks (ii-vi), notably absent peak i. By calculating the energy differences between these peaks, as detailed in Table I, we gain insight into the distinct spectral profiles of the LS and HS states. Remarkably, the energy discrepancies between peaks in both LS and HS phases are similar to those observed experimentally, with the HS phase showing a closer alignment. In our understanding, as the DFT+U energies of the LS and the HS are comparable within the reasonable range of U values (U=2-4eV), both the states can co-exist upon excitation. The absence of peak i in the LS phase suggests that the HS state may represent the initial (ground) state during the XAS excitation process.

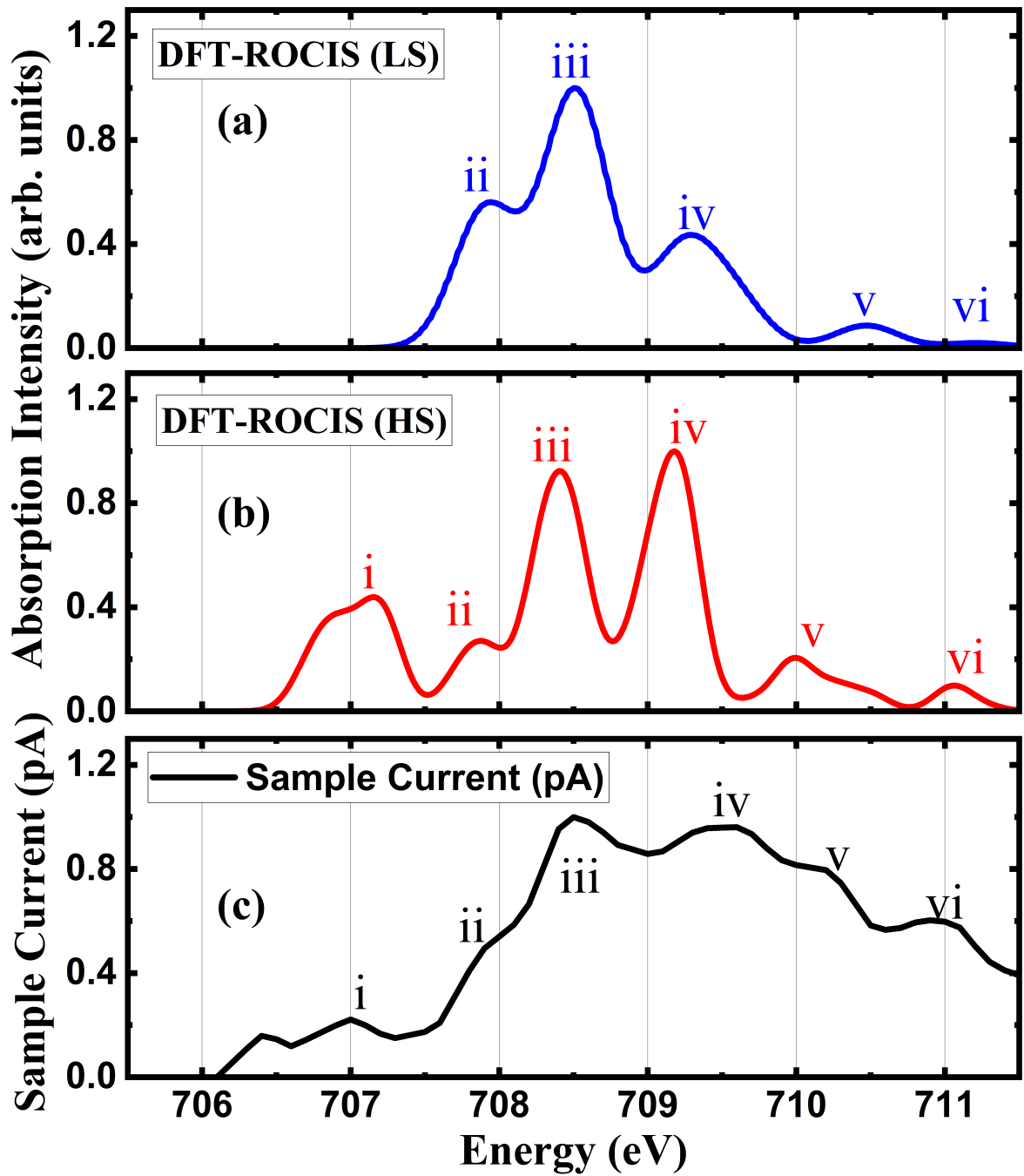

*Fig. 7.* Calculated L3-edge XAS spectra of $Fe^{2+}$ in $[Fe(tpy-Ph)_2]^{2+}$ for (a) LS and (b) HS states. (c) STM-NEXAFS signal of Fe ion experimentally measured.

**Table I:** Energy difference between peaks (from Figure 7) in XAS for LS and HS states.

| ΔE | Exp | LS | HS |
|---|---|---|---|
| i-ii | 1.0 | N/A | 0.9 |
| ii-iii | 0.7 | 0.6 | 0.6 |
| iii-iv | 0.6 | 0.8 | 0.8 |
| iv-v | 0.8 | 1.1 | 0.8 |
| v-vi | 1.1 | 0.8 | 1.1 |

We acknowledge that the TD-DFT methodology inherently assumes an initial state of excitation with U=0, reflecting a limitation in accounting for the electron correlation effects specifically related to the Fe d orbitals. Although the electron correlation in Fe-N bonding scenarios may be less pronounced than in Fe-based oxides—attributable to the stronger hybridization in Fe-N bonds compared to Fe-O bonds—a nonzero U parameter is anticipated for a more accurate representation. It is important to note, however, that our analysis indicates the electronic structures of each spin state remain largely consistent across varying U values. The distinction primarily lies in the separation of Fe d states, which is influenced by the Hubbard U parameter, suggesting that the orbital characteristics of each XAS peak are expected to be similar, even when a nonzero correlation U is factored in using more sophisticated approaches.

In addressing these findings, we conducted a comprehensive orbital analysis for each XAS peak, presented in Tables II and III. Table II explains the decomposed charge densities for the core hole Fe 2p states and the unoccupied Fe 3d states (where excitation of the core electron occurs) within the LS phase. It is observed that the core hole state demonstrates pronounced localization, whereas the Fe 3d states exhibit significant hybridization with the ligand. This hybridization is not limited to the immediate N atoms but also encompasses C atoms, indicating that a simplified Fe-N6 cluster approximation for XAS computations falls short of accurately depicting the complex Fe-ligand hybridization phenomena.

Upon analyzing the natural transition orbitals (NTOs) of the excited states in each prominent peak, it was observed that, for the LS phase (table II), peak ii is related to Fe $d_{x^2-y^2}$, iii related $d_{yz}$, iv related to $d_{z^2}$, v related to $d_{xy}$ and orbital, vi is related to $d_{z^2}$. The orbitals in these main peaks are also strongly hybridized with ligand N and C p-orbitals.

Similarly, NTOs for core hole and vacant states within the HS phase are shown in Table III. In this case, peak i is related to Fe $d_{xz}$, ii related to $d_{xy}$, iii related $d_{x^2-y^2}$, iv related to $d_{z^2}$, v related $d_{xz}$ and vi related to $d_{yz}$. In all these excited states, Fe 'd' states are hybridized with N 'p' states. Comparing the orbitals of LS and HS, stronger hybridization was observed in the case of the LS. From the NTOs analysis and comparing with the peak intensities, it is observed that the high intensity of peaks iii and iv correspond to the 'Fe' $d_{x^2-y^2}$ and $d_{z^2}$ states respectively. Surprisingly, it was observed that the orbitals in peak iv are similar in both LS and HS. Most interestingly, the orbitals in peak iv are oriented parallel to the plane. In our X-ray-excited resonance tunnelling (X-ERT) measurements, except peak iv (in which orbitals are aligned parallel to the surface), all other peaks were observed.[38] These findings help in understanding the nature of orbital excitations in each of the peaks that appear in SX-STM images. Thus, through DFT/XAS analysis, deeper investigation of the experimentally measured spectra can be done.

**Table II:** Natural transition orbitals of prominent states in each peak (ii to vi, refer to **Figure 7**) for low spin. Orbital excitation from 'Fe' core 2p orbital to all the main peaks unoccupied states are shown below. Column 2 details the corresponding excitation energies in eV.

| Band # (reference fig 6) | Energy (eV) | HOTO (core hole) | LUTO (electron/valence) |
|---|---|---|---|
| ii | *707.9* | 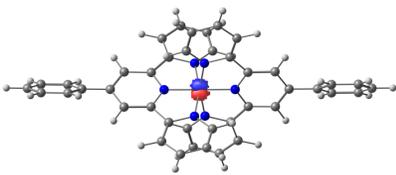 | 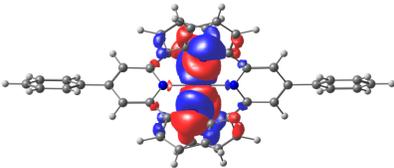 $N\,(p_z + p_y) + Fe\,d_{x^2-y^2}$ |
| iii | *708.5* | 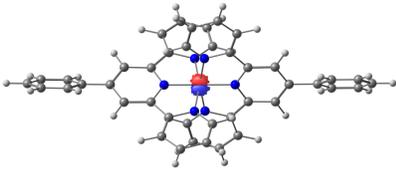 | 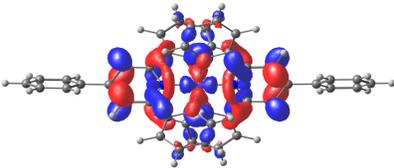 |

| | | | N $(p_x + p_y)$ + Fe $d_{yz}$ |
|---|---|---|---|
| iv | 709.3 | 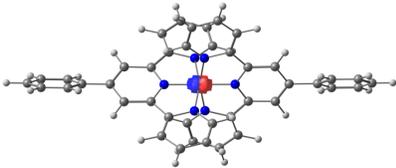 | 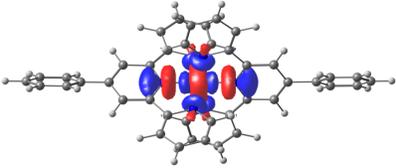<br>N $p_z$ + Fe $d_{z^2}$ |
| v | 710.5 | 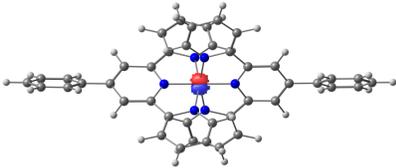 | 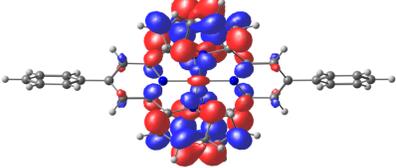<br>N $(p_x + p_y)$ + C p + Fe $d_{xy}$ |
| vi | 711.3 | 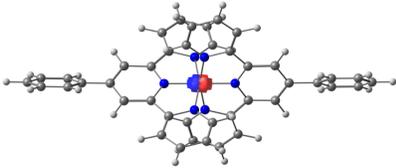 | 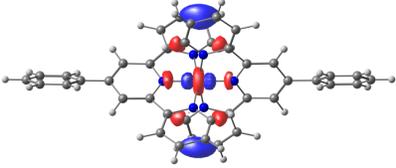<br>N $(p_x + p_y)$ + C p + Fe $d_{z^2}$ |

**Table III:** Natural transition orbitals of prominent states in each peak (ii to vi, refer Figure 7) for high spin. Orbital excitation from 'Fe' core 2p orbital to all unoccupied states are shown. Column 2 details the corresponding excitation energies in eV.

| Band # (reference fig 6) | Energy (eV) | HOTO (core hole) | LUTO (electron/valence) |
|---|---|---|---|
| i | *707.0* | 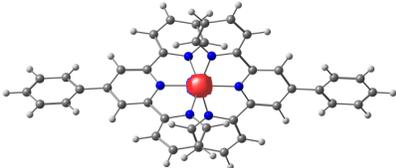<br>$p_x$ | 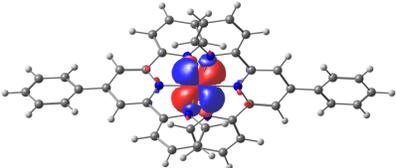<br>N $p_x$ + Fe $d_{xz}$ |
| ii | *707.9* | 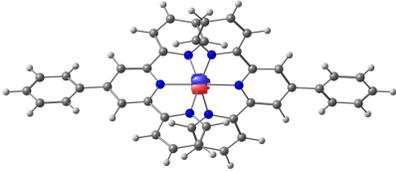<br>$p_y$ | 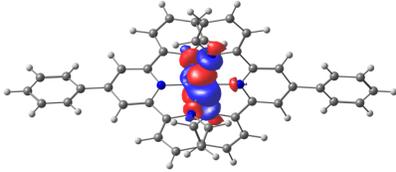<br>N $(p_z + p_y)$ + Fe $d_{xy}$ |
| iii | *708.5* | 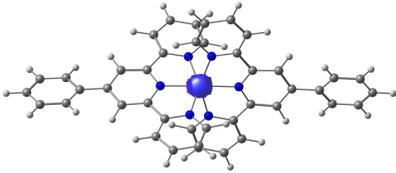<br>$p_x$ | 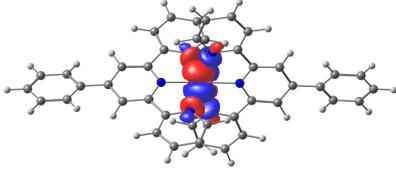<br>N $p_y$ + Fe $d_{x^2-y^2}$ |
| iv | *709.3* | 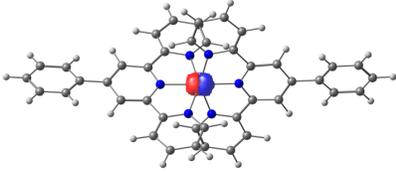<br>$p_z$ | 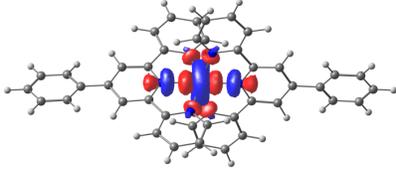<br>N $p_z$ + Fe $d_{z^2}$ |
| V | *710.1* | 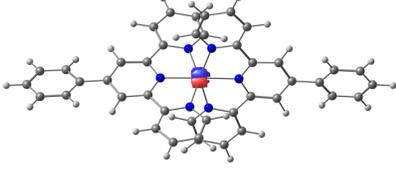<br>$p_y$ | 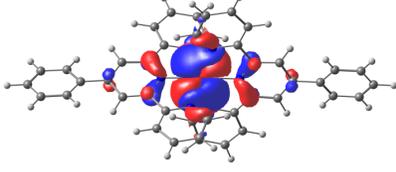<br>N $(p_x + p_y)$ + Fe $d_{xz}$ |
| vi | *711.2* | 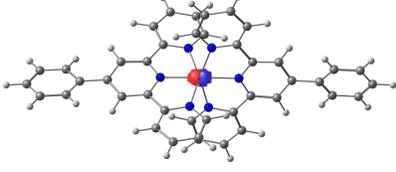<br>$p_z$ | 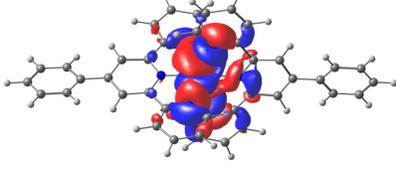<br>N $(p_z + p_y)$ + Fe $d_{yz}$ |

**Conclusions:**

In summary, using density functional theory + U (DFT+U) calculations, we have detailed the electronic nuances of the $Fe^{2+}$ ion in both its low-spin (S=0, $t_{2g}^6 e_g^0$) and high-spin (S=2, $t_{2g}^4 e_g^2$) states. Our findings reveal that the low-spin state is energetically favorable at U(Fe) values of 3 eV or lower, whereas the high-spin state becomes stable at U(Fe) values above 3 eV for the free *[Fe(tpy-Ph)₂]²⁺* molecule. The interaction between *[Fe(tpy-Ph)₂]²⁺* and Au substrate, characterized as predominantly van der Waals, does not significantly alter the Fe d electronic structure but slightly modifies the critical U value for the spin-state transition to approximately 3.5 eV.

The application of time-dependent DFT (TD-DFT) for calculating the L-edge X-ray absorption spectra (XAS) of both spin states, and correlating these with SX-STM experimental data, has further refined our understanding of *[Fe(tpy-Ph)₂]²⁺*. The XAS data from the high-spin phase align more closely with experimental observations, suggesting the high-spin state might exist as the initial state during the X-ray excitation process. Such insights enrich our understanding of the complex spin state interactions within *[Fe(tpy-Ph)₂]²⁺*, emphasizing the importance of recognizing all possible spin states to fully appreciate the molecular behavior and dynamics, and spotlighting the pivotal roles of both high-spin and low-spin states in these processes.

# Appendix

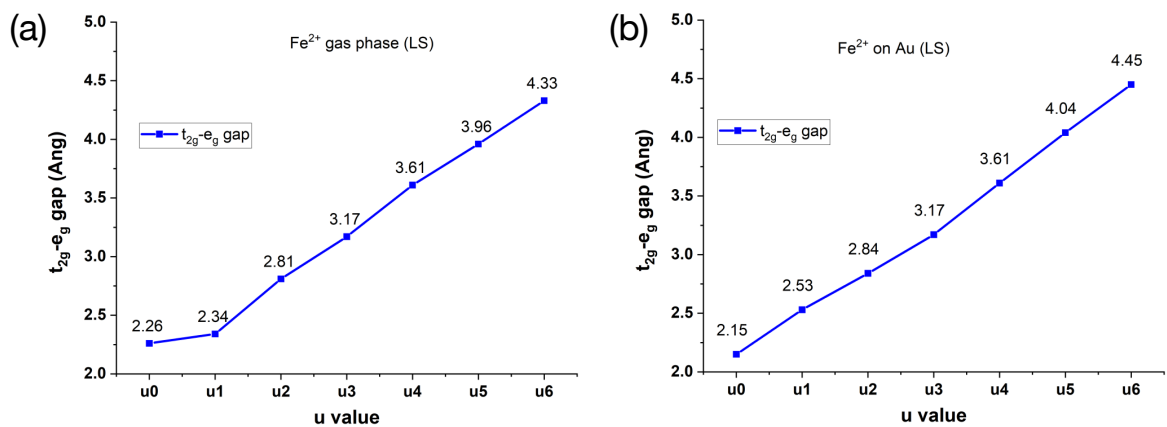

**Fig 5** *of the LS state as a function of U for [Fe(tpy-ph)₂]²⁺*